\documentclass{webofc}
\usepackage[varg]{txfonts}   
\usepackage{tikz}
\usetikzlibrary{decorations.pathmorphing}
\usetikzlibrary{arrows.meta}
\tikzset{snake it/.style={decorate, decoration=snake}}

\begin{document}
\title{Transfer Reactions in Nuclear Astrophysics}
%
%

\author{\firstname{Philip} \lastname{Adsley}\inst{1,2}\fnsep\thanks{\email{padsley@tamu.edu}} 
}

\institute{Cyclotron Institute, Texas A\&M University, College Station, Texas 77843-3366, USA
\and
 Department of Physics and Astronomy, Texas A\&M University, College Station, Texas 77843-4242, USA
          }

\abstract{%
  Transfer reactions are important tool in nuclear astrophysics. These reactions allow us to identify states in nuclei and to find the corresponding energies, to determine if these states can contribute to astrophysical nuclear reactions and ultimately to determine the strength of that contribution. In this paper, the basic details of how transfer reactions may be used in nuclear astrophysics are set out along with some common pitfalls to avoid.
}
%
\maketitle
\section{Introduction}
\label{intro}

Direct measurements of nuclear cross sections at energies relevant to stellar burning are extremely challenging. Indirect methods provide an alternative way of determining astrophysically important reaction rate. Transfer reactions are one of these indirect methods. This paper briefly sets out the background of transfer reactions and their utility in nuclear astrophysics. More detailed discussions are available in other sources such as Refs. \cite{IliadisBook,FH_NdS_paper}.

\begin{figure}[htbp]
    \centering
    \begin{tikzpicture}
    \filldraw(0,0) circle (0.25cm);
    \draw(0,-0.25) circle(0.25cm);
    \draw[dotted] (0,-0.125) ellipse (0.35 cm and 0.55 cm);
    \node at (-0.5,0) {$n$};
    \node at (-0.5,-0.25) {$p$};
    
    \draw(2,0.5) circle(0.5cm);
    \node at (3,0.5) {$^{20}$Ne};
    
    \draw(6,0.5) circle (0.5cm);
    \filldraw(6,0.0) circle (0.25cm);
    \draw[dotted] (6,0.35) ellipse (0.55 cm and 0.85 cm);
    \node at (7,0.35) {$^{21}$Ne};

    \draw(8,-0.5) circle (0.25cm);
    \node at (8,-1.0) {$p$};
    \draw[-latex] (8+0.25*0.97,-0.5-0.25*0.2588) -- (8+1.5*0.97,-0.5-1.5*0.2588);
    
    \draw[-,dotted] (4,-2) -- (4,2);

    \node at (1,2) {Before};
    \node at (7,2) {After};
    \end{tikzpicture}
    \caption{A schematic diagram showing a transfer reaction, in this case $^{20}$Ne($d,p$)$^{21}$Ne. The deuteron projectile is shown as a neutron (black filled circle) and a proton (small hollow circle) bound together, and the $^{20}$Ne target is shown as a large hollow circle. }
    \label{fig:SchematicTransfer}
\end{figure}
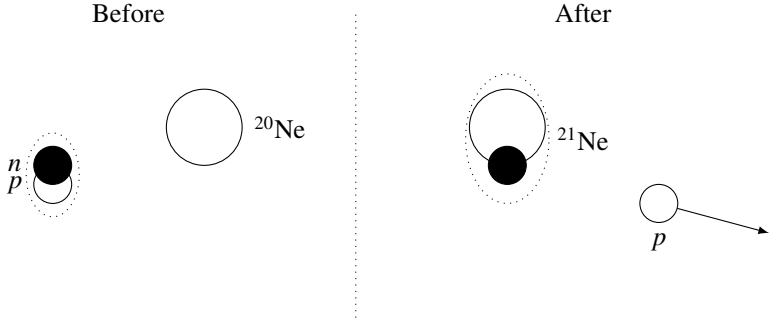

Transfer reactions are, unsurprisingly, reactions in which something is transferred. This transferred item can be a single nucleon (a proton or a neutron), a cluster (such as an $\alpha$ particle or a deuteron) or can include swapping two particles in charge-exchange reactions such as $(p,n)$ or $(^3\mathrm{He},t)$. Transfer reactions happen quickly, typically on the order of $10^{-22}$ seconds, and are dominated by peripheral reactions where the nuclear reaction takes place at the surface of the nucleus. A schematic diagram for a transfer reaction is shown in Fig. \ref{fig:SchematicTransfer}.

In general, to model transfer reactions we used calculations such as the (first-order) Distorted-Wave Born Approximation (DWBA). More complex reaction models such as coupled-channel calculations are sometimes necessary but beyond the current scope; for a comprehensive treatment, see the book by Thompson and Nunes \cite{TandN}. These calculations themselves require a number of inputs, including experimental values such as the beam and target species, the beam energy, the spins of the various states, the transferred spin and orbital angular momentum etc. There are also other inputs such as optical-model potentials. These potentials describe the interaction between the nuclei in the entrance channel and the nuclei in the exit channel, the binding of transfer fragment to the core systems in both the initial and final partitions, and possibly other interactions such as the core-core interaction.

\section{Background}
\label{sec:BG}

In order to calculate thermonuclear reaction rates in stars the cross section, $\sigma(E\mathrm{c.m.})$, which represents the probability of interaction between two particles with a centre-of-mass energy $E_\mathrm{c.m.}$, must be folded with the Maxwell-Boltzmann distribution, representing the likelihood of the two particles having that relative energy. The thermonuclear reaction rate is then found by integrating this product across all energies, representing a sum over all possible energies of the product of the interaction probability with the probability of two particles having that relative energy.

The cross section can have a number of different contributions including direct-capture and resonance reactions. Direct capture is an electromagnetic transition from an initial continuum state in which a projectile is moving outside a nucleus to a final bound state within the nucleus. Since the electromagnetic force is well known, the inputs from nuclear physics are the energy, spin and parity of the bound state, and the overlap between the initial nucleus+projectile configuration and the final bound state in the new residual nucleus. This later quantity is expressed as the spectroscopic factor, $C^2S$, or the asymptotic normalisation coefficient. The direct-capture cross section can be calculated using $\sigma_\mathrm{DC}(E_\mathrm{c.m.}) = \sum_i (C^2S)_i\sigma_{i,\mathrm{DC}}(E_\mathrm{c.m.})$, where $(C^2S)_i$ is the spectroscopic factor of the $i$th state and $\sigma_{i,\mathrm{DC}}(E_\mathrm{c.m.})$ is the single-particle direct-capture cross section at the centre-of-mass energy $E$ which is calculated analytically using codes such as TEDCA \cite{FH_NdS_paper,TEDCA}.

Resonance reactions are two-step processes - first the incoming projectile fuses with the target nucleus, forming an excited resonance state. Next, this resonance state decays, usually by the emission of another particle or a $\gamma$ ray. For resonance reactions, the cross section can vary significantly more than for direct capture. The reaction cross section is typically written as in Breit-Wigner form:
\begin{equation}
    \sigma_\mathrm{res}(E_\mathrm{c.m.}) = \omega \frac{\pi\hbar^2}{2\mu E}\frac{\Gamma_\mathrm{in}\Gamma_\mathrm{out}}{(E_\mathrm{c.m.}-E_R)^2 + \Gamma^2/4}
\end{equation}
where $\omega = \frac{2J+1}{(2j_1+1)(2j_2+1)}$ is the spin-statistical factor with $J$ the spin of the resonance and $j_{1,2}$ the spins of the reactants, $\mu$ the reduced mass, $\Gamma_{in(out)}$ the partial width for the entrance (exit) channel, $\Gamma$ is the total width $\Gamma = \sum_i \Gamma_i$, $E_R$ is the resonance energy and $E_\mathrm{c.m.}$ is the centre-of-mass energy at which the reaction takes place.

\begin{figure}[htbp]
\centering
\begin{tikzpicture}
\draw[-] (-2,0) -- (0,0);
\draw[-{Latex[scale=1.5]},red] (-0.5,0) -- (-0.5,2);
\draw[-,dotted] (-0.5,2) -- (1.5,2);
\draw[-,red] (1.5,2) -- (3.5,2);
\draw[-] (1.5,-4) -- (3.5,-4);
\draw[-{Latex[scale=1.5]},red, snake it] (2.0,2) -- (2.0,-4);
\draw[-{Latex[scale=1.5]},snake it] (-0.5,0) -- (1.8,-4);
\draw[-,dotted] (0,0) -- (3.5,0);
\draw[{Latex[scale=1.5]}-{Latex[scale=1.5]},red] (3.0,-4) -- (3.0,2);

\node at (2.5,-4.75) {$^{A+1}Z+1$};
\node at (-1,-0.75) {$^AZ$};
\node at (-1,1) {\textcolor{red}{$E_\mathrm{c.m.}$}};
\node at (3.75,0) {$S_p$};
\node at (4.25,-2) {\textcolor{red}{$E_\mathrm{x} = S_p + E_\mathrm{c.m.}$}};
\end{tikzpicture}

\caption{Schematic diagrams showing the (black) direct-capture mechanism and (red) resonance-capture mechanism for proton capture onto a target nucleus $^AZ$ to the ground state of the residual nucleus, $^{A+1}Z+1$. The excitation energy of the resonance state of importance for the reaction is given by the sum of the proton separation energy, $S_p$, and the centre-of-mass energy of the colliding system, $E_\mathrm{c.m.}$.}
\label{fig:DC}       
\end{figure}
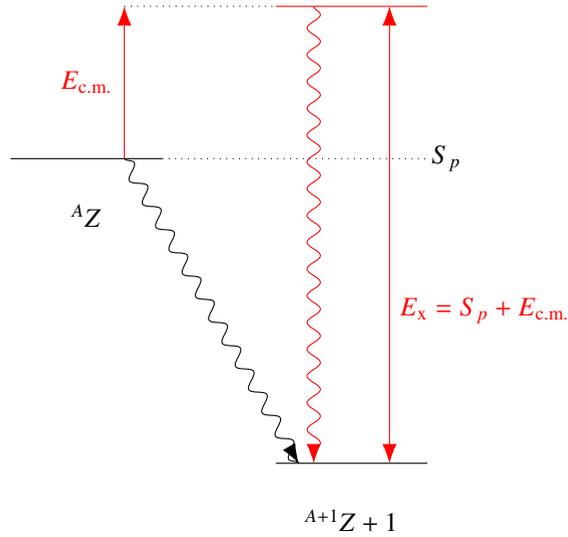

\section{Uses of transfer reactions in nuclear astrophysics}

Having covered the background of how nuclei react through different mechanisms, the use of transfer reactions in determining the relevant nuclear quantities of interest is now discussed. From the previous section, it is possible to list the nuclear quantities required to calculate the reaction rate. These include the existence and energies of nuclear states, their spin and parity, and their overlap with various initial and final configurations including various different quantities defined below such as the spectroscopic factor, the asymptotic normalisation coefficient, and the partial width.

One important consideration when using transfer reactions for nuclear astrophysics is that the information gleaned is model-dependent and, if inappropriate models are used to describe the data, then the conclusions drawn are potentially deficient. 

\subsection{State and resonance energies}

The simplest use of transfer reactions is in finding if states exist and, if so, at what energies they occur in the compound nucleus. The energies of states are determined by using two-body kinematics for the transfer reaction and $E_{x}/E_{r}$ reconstructed from the beam energy, the known masses and the detected energies and angles of the emitted radiations.

One complication is that transfer reactions tend to be selective; this is sometimes a great advantage but it can be a problem. Many transfer reactions are selective to the states populated and the selectivity may not be to the states which are most important astrophysically. This can lead to bias since only a subset of the states are used to evaluate the reaction rate. Some caution is required in drawing firm conclusions as to the non-existence of states from only a single transfer reaction.

\begin{figure}
    \centering
    \includegraphics[width=0.75\textwidth]{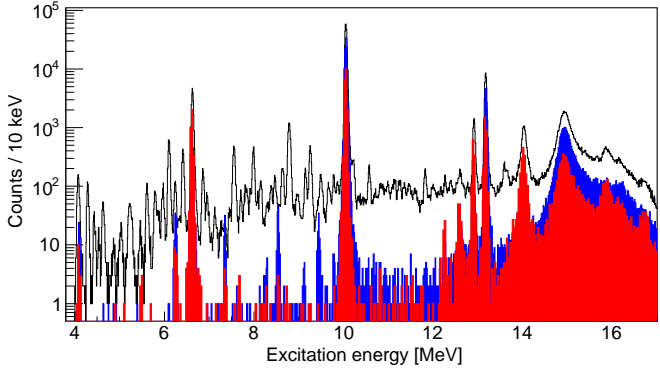}
    \caption{An excitation-energy spectrum of states in $^{48}$Cr following the $^{50}$Cr($p,t$)$^{48}$Cr reaction \cite{Sifundo}. The hollow black spectrum corresponds to states in $^{48}$Cr while the blue and red spectra show $^{12}$C($p,t$)$^{10}$C and $^{16}$O($p,t$)$^{14}$O reactions from carbon and Mylar targets, respectively.}
    \label{fig:Sifundo_50Cr_pt}
\end{figure}

An example of a study in which a transfer reaction has been used to find excited states in a nucleus is $^{50}$Cr($p,t$)$^{48}$Cr, measured a number of years ago with the K600 magnetic spectrometer and an array of silicon detectors \cite{K600,CAKE,Sifundo} at iThemba LABS, South Africa. Many states in $^{48}$Cr were observed for the first time, states which have a potential impact on the $^{44}$Ti($\alpha,p$)$^{47}$V reaction which destroys $^{44}$Ti produced in core-collapse supernovae. An excitation-energy spectrum for this reaction is shown in Fig. \ref{fig:Sifundo_50Cr_pt} showing the large number of $^{48}$Cr states populated, many observed for the first time, giving valuable information on the accuracy of calculations of the $^{44}$Ti($\alpha,p$)$^{47}$V reaction rates.

\subsection{Spins and parities}

Spins and parities can be constrained using transfer reactions. Spins and parities are needed because they determine which possible orbital angular momenta can contribute to astrophysical reactions. An higher orbital angular momentum means a lower penetrability through the barrier for the incoming particle, and a smaller contribution to the reaction rate. From the differential cross section, the probability of having a reaction take place at a particular angle, the transferred angular momentum of a reaction can be determined.

Again, there are complications with this process. One is that many transfer reactions are insensitive to the total spin of the final state, with the gross structures of the differential cross section being determined predominantly from the transferred angular momentum. This means that, for many single-nucleon transfer reactions, the spin of the final state can only be determined to $j = \ell\pm1/2$. There are some exceptions to this: first, the spin-orbit term of the potential can induce some fine structure which can allow for discrimination between various final spins and, second, transfer reactions in which tensor analysing powers can be measured (e.g., using polarised deuterons in the $\left(\overrightarrow{d},p\right)$ reaction) allows for discrimination between different final spins.

\subsection{Spectroscopic factors and asymptotic normalisation coefficients}

Both direct-capture and resonance-capture reactions rely on the behaviour of the wave function of the final state. For direct capture, what matters is how like the initial configuration the final state looks, described by the spectroscopic factor. For resonance capture, what matters is how likely the transferred particle is to appear at the nuclear surface in the final state; this, in turn, depends on the single-particle wave function and the spectroscopic factor. The likelihood of appearance is then multiplied by the likelihood of the transferred particle making it through the potential barrier for the nucleus (the penetrability) to get the partial width for the state. Due to time-reversal invariance, this partial width is also the likelihood of getting that particle into the nucleus through the potential barrier into the resonance state in question, as needed to calculate the cross section.

The spectroscopic factor, $C^2S$ is an empirical factor which relates the calculated cross section to the observed cross section $\frac{d\sigma}{d\Omega}_\mathrm{exp} = C^2S \frac{d\sigma}{d\Omega}_\mathrm{calc}$ where $\frac{d\sigma}{d\Omega}_\mathrm{exp(calc)}$ is the experimental (calculated) differential cross section.

Spectroscopic factors can be used to calculate the direct-capture reaction component as described in Section \ref{sec:BG}. For resonance-capture reactions, the quantity of interest is the partial width. The partial width can be written as \cite{IliadisBook}:
\begin{equation}
    \Gamma = 2P_\ell(E,R)\frac{\hbar^2R}{2\mu}C^2S|\phi(R)|^2,
\end{equation}
where $P_\ell$ is the penetrability through the barrier with angular momentum $\ell$, $\mu$ is the reduced mass, and $\phi(R)$ is the radial wave function desribing the relative motion of the transferred nucleon in the final state at the radius $R$. The spectroscopic factor here ($C^2S$) is the same as previously. However, the measured spectroscopic factor depends on the potentials chosen to model the reaction. For nuclear astrophysics, it is better to calculate the partial width using the same potentials and radial wave functions as the DWBA calculation and determination of the spectroscopic factor. This will not remove the entire model-dependency of the reaction but the uncertainties in the partial width from the reaction model are significantly reduced by performing self-consistent calculations, see Ref. \cite{Harrouz} for an example.

The asymptotic normalisation coefficient (ANC) measures the normalisation of the tail of the nuclear overlap, and are used both in the calculation of direct-capture reactions into bound states and the determination of the partial width for resonances \cite{TandN}. ANCs are less model-dependent in terms of the connection between the measured transfer cross section and the partial width of interest. However, since ANCs are usually extracted from sub-Coulomb transfer reactions which are extremely sensitive to the energy of the interaction and have a small cross section, the extracted partial widths often have significant uncertainties. A good example of the difficulty of measuring ANCs is given by the factor 3 change in the ANC for $^{13}$C$+\alpha$ between the works of Johnson {\it et al.} \cite{Johnson} and Avila {\it et al.} \cite{Avila}, the latter study demonstrating that the former had an uncorrected problem due to build-up on the target reducing the effective beam energy. These extremely experimentally challenging studies are vital for nuclear astrophysics but require considerable expertise.

Transfer reactions typically only provide information on one part of a resonance reaction. However, since the cross section is typically dominated by the slowest step (i.e. the smallest partial width), the information obtained from the transfer reaction may not entirely specify the reaction rate. An excellent example of this was the $^{25}$Mg($d,p$)$^{26}$Mg study of Chen {\it et al.} \cite{Chen} and its comparison to $^{25}$Mg$+n$ measurements \cite{Massimi1,Massimi2}. Since the $\gamma$-ray partial width is generally smaller for most of the relevant states, the $^{25}$Mg($n,\gamma$)$^{26}$Mg cross section is governed by the $\gamma$-ray partial widths. Neutron widths can be determined from ($d,p$) reactions but these may not tell the whole story. Coincidence measurements of the ($d,p\gamma$) reaction may be necessary. For many reactions, especially charged-particle reactions, the particle partial width is the smallest and therefore rate-limiting width which means that the reaction rate can be totally specified from the transfer reaction.

\section{Summary}

Transfer reactions remain one of the most powerful tools for determining thermonuclear reaction rates for nuclear astrophysics, having been a workhorse of the field for decades. Transfer reactions can be used to locate astrophysically important states and measure their properties so that astrophysical reaction rates may be calculated. As with every experimental technique, the devil is in the details. In this paper, the broad background of transfer reactions has been laid out along with some potential pitfalls and problems.

%

\begin{thebibliography}{}
%
%
\bibitem{IliadisBook}
C. Iliadis, Nuclear Physics of Stars, Wiley-VCH, Weinheim (2007)
\bibitem{FH_NdS_paper}
F. Hammache and N de S\'{e}r\'{e}ville, Frontiers in Physics \textbf{8} 602920 (2021)
\bibitem{TandN}
I. Thompson and F. Nunes, Nuclear Reactions for Astrophysics, Cambridge University Press, Cambridge (2009).
\bibitem{TEDCA}
H. Krauss, TEDCA, unpublished computer code (1992)
\bibitem{CAKE}
P. Adsley {\it et al.}, Journal of Instrumentation \textbf{12}, T02004 (2017).
\bibitem{K600}
R. Neveling {\it et al.}, Nuclear Instruments and Methods in Physics A \textbf{654}, 29-39 (2011).
\bibitem{Sifundo}
S. D. Binda, MSc Dissertation, University of the Witwatersrand, Johannesburg, South Africa, (2022); S. D. Binda {\it et al.}, paper in preparation.
\bibitem{Johnson}
E. D. Johnson {\it et al.}, Physical Review Letters \textbf{97}, 192701 (2006).
\bibitem{Avila}
M. L. Avila {\it et al.}, Physical Review C \textbf{90}, 024327 (2014).
\bibitem{Harrouz}
D.S. Harrouz {\it et al.}, Physical Review C \textbf{105}, 015805 (2022).
\bibitem{Chen}
Y. Chen {\it et al.} Physical Review C \textbf{103} 035809 (2021)
\bibitem{Massimi1}
C. Massimi {\it et al.} Physics Letters B \textbf{768} 1 (2017)
\bibitem{Massimi2}
C. Massimi {\it et al.} Physical Review C \textbf{85} 044615 (2012)
\end{thebibliography}
%
%

\end{document}